# Origin and Electronic Behavior of Improper Ferroelectricity in AB2 (A=Cr, Mo, W; B=S, Se, Te) Transition Metal Dichalcogenides

Chengcheng Xiao[1,2], Zhu-An Xu[2], Xin Luo[3*], Yunhao Lu[2,4*]

[1] Departments of Materials and Physics, and the Thomas Young Centre for Theory and Simulation of Materials, Imperial College London, London SW7 2 AZ, UK

[2] Zhejiang Province Key Laboratory of Quantum Technology and Device, Department of Physics, Zhejiang University, Hangzhou, 310027, China

[3] School of Physics, Sun Yat-sen University, Guangzhou, 510275, China

[4] State Key Laboratory of Silicon Materials, School of Materials Science and Engineering, Zhejiang University, Hangzhou, 310027, China

Correspondence and requests for materials should be addressed to X. Luo (luox77@mail.sysu.edu.cn) & Y.H.Lu (luyh@zju.edu.cn)

(Date: Jan 03, 2020)

## Abstract

Persistent electrical polarized states are fundamentally important to the electric industry as they can be used in the non-volatile memory, the artificial neuromorphic network, and negative capacitors, making ultralow energy consumption electronic devises possible. With the recent development in low dimensional ferroelectric materials, emerging 2D out-of-plane ferroelectric materials like $MoTe_2$ have great potential for future development. Despite previous phenomenological studies, the underlying microscopic origin of ferroelectricity is still missing. Here, using density functional theory and Wannier function methods, we reveal that the origin of ferroelectricity of these transition metal dichalcogenides comes from the Jahn taller effect and followed by a covalent bonding between transition metal atoms. Moreover, the atypical electronic behavior of these monolayer $AB_2$ (A=Cr, Mo, W; B=S, Se, Te) TMDs compare to traditional improper ferroelectrics hints a strong electronic origin of the polarization, suitable for future industrial applications.

Ferroelectricity possess the potential for numerous devices and memory applications due to its switchable spontaneous electrical polarizations. The prototypical perovskite ferroelectrics like $BaTiO_3$ and $PbTiO_3$ have been widely studied. The underlying mechanism of structural phase transition in perovskite ferroelectrics is revealed as partial covalent contributions to the ionic bonding, thus stabilizing the distorted non-centrosymmetric ferroelectric phase[1, 2]. However, when thinning down to few layers, the polarization of perovskite material vanishes due to strong depolarization field and the breaking of long range dipole coupling. Recently, the discovery of two-dimensional ferroelectric materials such as SnTe[3] and SnSe[4] have sparked a new round of research for this phenomenon in low-dimension materials. The majority of proposed materials like Group-V elemental monolayers[5] and VI-IV compound monolayers[6, 7] possess in-plane polarizations while several out-of-plane monolayer ferroelectric materials such as $In_2Se_3$[8], $CuInP_2S_6$[9, 10] and $MoS_2$[11] have also been theoretically proposed. The most peculiar one of which is the out-of-plane ferroelectricity reside in the monolayer $MoS_2$ revealed by Sharmila et al.[11]. It is the first time the out-of-plane polarization have been predicted theoretically and subsequently found in experiments in a similar $MoTe_2$ monolayer[12]. Most literature studying the ferroelectric properties of these $AB_2$ (A=Cr, Mo or W, B=S, Se, Te) transition metal dichalcogenides (TMD) systems rely on phenomenological or statistical analysis while trying to discern the reason behind the breaking of so called "scaling law" of ferroelectricity[13]. However, a microscopic explanation for the origin of the ferroelectricity is still lacking. Albeit phenomenological study can give a fairly consistent prediction of structural transition comparing to the experiments, a microscopic study has the ability to closely examine the electronic and structural relations and can be beneficial for designing a general process of such improper ferroelectric materials.

In this paper, with the help of First-principle calculations, we studied the structural phase transition and its electronic origin of monolayer $AB_2$ structures on a microscopic level. Since the out-of-plane polarization of monolayer $MoTe_2$ has already be confirmed by experiment and other AB$_2$ structures possess similar properties [Fig. S1], later sections will mainly focus on the MoTe$_2$ monolayers. As seen from below discussion, the polarization of MoTe$_2$ monolayer largely originates from the primary distortions of Mo trimerization. This differ from the behavior of typical improper ferroelectrics where the second order polar mode usually generates the electrical polarization[14, 15]. Our results based on Wannier interpolation of Bloch bands and Wannier orbital overlap populations (WOOP) method[1, 16] reveal a fairly strong covalent bonding which causes the transition between different phases in monolayer MoTe$_2$. Combined with a detailed symmetry analysis, we were able to simulate the phase response to the external electric field, enhancing the understanding of the origin and the stability of this "electronic" improper ferroelectricity in d1T AB$_2$ monolayers.

All first-principles calculations are carried out within the density functional theory as implemented in the Vienna Ab-initio Simulation Package (VASP) software[17]. The Local-Density-Approximation (LDA) combined with a projector augmented-wave pseudopotential[18] was employed and the Mo's *4s 5s 4p 4d* and Te's *5s 5p* electrons were treated as valence electrons. For electronic structure calculations, plan-waves with kinetic energy above 500eV was truncated and an energy convergence criterial of 1E-8 eV was employed. All calculations were performed within the non-relativistic approximation, excluding the spin-orbit coupling effect. To avoid interaction between different imaging structure due to periodic boundary condition and electrical polarization, a vacuum space of around 30 Å was added along with a dipole correction layer inside the vacuum space. A 12×12×1 Γ centered Monkhorst-Pack[19] grid were chosen for the K-space sampling of d1T structure and are adjusted proportionally according to the lattice parameters of different structures. In geometry optimizations, the ionic positions were sufficiently relaxed until all forces acted on every atom were less than $10^{-3}$ eV/Å. The obtained lattice parameters as well as proportional atomic positions can be found in the supporting documents[20]. Wannier interpolation of Bloch bands were carried out using the WANNIER90 package[21]. The

Wannier functions were generated by projecting the Bloch wave function on to a set of Hydrogen like atomic orbitals. Wanneir Orbital Overlap Populations were calculated using the WOOPS package[22], detailed descriptions of this method can be found in the supporting documents[20]. The symmetry analysis and the construction of Landau polynomials are done using the ISOTROPY suite[23].

There are several polymorph forms of $AB_2$ monolayers while the H phase being widely accepted as the most stable one[24]. Although without inversion centers, the H phase of $AB_2$ monolayers have a space group of $P\bar{6}m2$ which possess a three-fold rotation axis, preventing the existence of ferroelectricity. Comparing to the H phase, the energetically lesser stable 1T phase of these $AB_2$ monolayers have a centro-symmetric space group of $P\bar{3}m1$, also prohibiting the emergence of ferroelectricity. However, due to the intrinsic dynamical instability, the 1T phase has the potential of distorting into other phases with hidden out-of-plane ferroelectric polarizations. The 1T phase, taking the $MoTe_2$ as example, possesses three atomic planes: the upper Te atom plane, the Mo atom plane and the lower Te atom plane [Fig. 1]. Each Mo atom in the 1T structure is surrounded by 8 Te atoms forming an equal lateral octahedron as shown in Fig. 1 and Fig. 3(b). The inversion center is located on one of the Mo atoms' atomic sites. As proposed by Sharmila et al.[11], by moving the Mo atoms positions in a specific pattern in this 1T structure, the out-of-plane symmetry can be subsequently broken, allowing the existence of ferroelectricity.

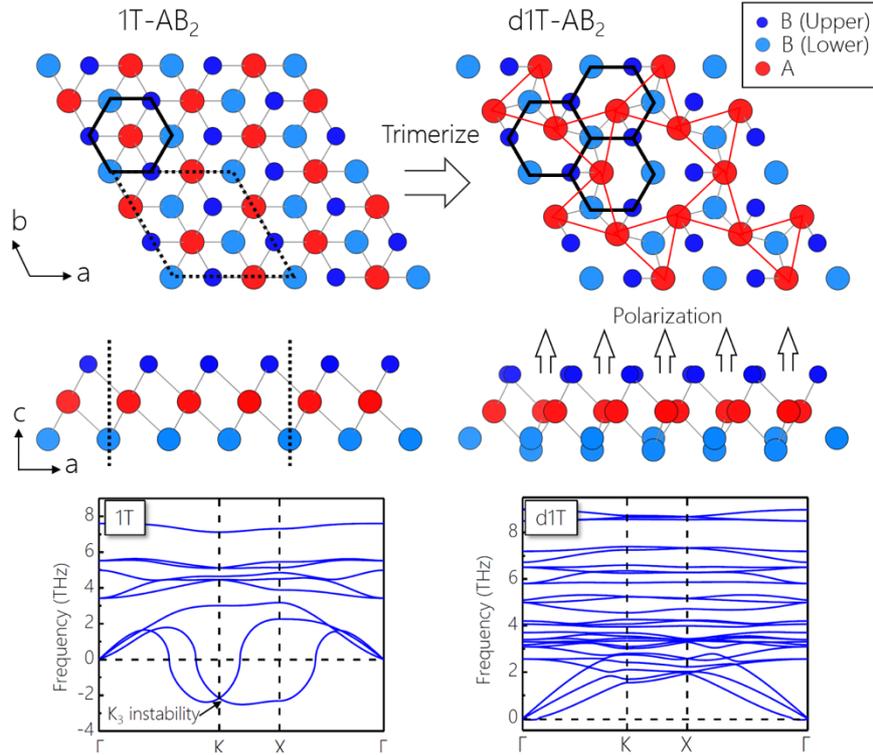

FIG. 1 Upper panel: Top view and side view of $AB_2$ transition metal dichalcogenide monolayers. Lower panel: Phonon dispersion of centrosymmetric 1T and d1T structure of $MoTe_2$ monolayer. The so-called $K_3$ instability is located at K point in the Brillouin zone of the 1T structure while the d1T structure shows no significant imaginary frequency.

To study the structural phase transitions of ferroelectrics, a phonon analysis of the centro-symmetric 1T phase was performed. The phonon analysis is very helpful since not only can it show the dynamical stability of the structure, but also to suggest potential displacement of atoms and many other properties like the type of proper or improper ferroelectricity. The phonon dispersion of centro-symmetric 1T $MoTe_2$ are shown in the lower panel of Fig. 1. The most pronounced instabilities are located slightly deviated from the K high-symmetry point, where on the K point, two instable(imaginary) optical phonon branch become energetically degenerate. Considering the symmetry related K' point, the total degeneracy of these unstable modes on K(K') point is 4. Comparing to the Γ point soft mode in traditional perovskite materials, these soft modes in Brillouin zone edge suggest a supercell is needed in order to stabilize our structure. By modulating the 1T $MoTe_2$ structure with these degenerate phonon modes' eigenvectors in a $\sqrt{3}$ supercell, a distorted 1T phase (d1T) can be constructed. As shown in the right panel of Fig.1, the d1T structure have a space group of $P31m$ which consists a trimerization movement of Mo atoms. Switching to the top view in Fig. 1, the trimerization of

Mo atoms happens around Te atoms located at the center of the Mo's equilateral triangles (connected with red lines). Taking a closer look at the d1T structure, the atomic sites can now be categorized by four sublattices: Upper Te position, Middle Mo position and two different lower Te positions. Comparing to the 1T structure, the lower Te layer now have two different sub-lattices with different positions in the out-of-plane c axis which is entirely different from the upper Te layer, effectively breaking the out-of-plane symmetry. With trimerization, the out-of-plane symmetry was broken accompanied by the rise of dipole moments, forming electronic polarization. The phonon dispersion of this new structure shown in the lower panel of Fig. 1 confirms that it is indeed dynamically stable. Meanwhile, the "flat" soft optical mode of the 1T phase from K to X point resemble that of a typical order-disorder ferroelectric material. Compare to the order-disorder limit, the soft optical branch here only exist at the edge of Brillouin zone while the order-disorder ferroelectrics have soft mode across entire Brillouin zone. This phenomenon suggests the coupling between neighboring Mo atoms' displacement can be relatively weak hence the prototypical 1T phase can be distorted into other different structures while the directions of the potential distortions are constrained. Actually, another distorted structure with an inversion symmetric $P21/m$ space group called 1T' structure can be constructed [Fig. S2] using soft mode at X point. Recently, it is reported that by stacking multiple 1T' layers together, the inversion symmetry can be broken due to the stacking sequence and out-of-plane ferroelectric polarization are measured in a few layers, and even in bulk system [25,26]. The multilayer or bulk structures are beyond the scope of this study and will not be discussed here. With the distorted d1T phase of $MoTe_2$ monolayers carefully defined, we can now proceed to phenomenological analysis of such phase and its relation/transition between 1T and distorted structures.

The phase space of a ferroelectric material is of most importance since the thermal excitation can break the ordered ferroelectric phase, rendering the system to a non-polarized paraelectric (PE) state. Fully mapping the potential surface of this material is a computationally formidable task, the total degrees of freedom can be as large as 27 in a $\sqrt{3}$ supercell. Luckily, with the help of space group symmetry, we were able to lower the degree of freedom drastically. To represent the full phase space and to allow all potential routes for the ferroelectric switching, we choose the lowest symmetry space group to be $P1$ while using the 1T phase ($P\bar{3}m1$) as the reference structure. The total free energy was expanded as a Tylor series consists of terms of coupling between different displacement of atoms called order parameters. The resulting polynomial takes the following form:

$$\begin{aligned}\mathcal{F}(n_1, n_2, n_3, n_4, n_5, n_6) \\
= g_{11}n_6 + g_{21}n_6{}^2 + g_{22}(n_1{}^2 + n_2{}^2 + n_3{}^2 + n_4{}^2) + g_{23}n_5{}^2 + g_{31}n_6{}^3 \\
+ g_{32}n_6(n_1{}^2 + n_2{}^2 + n_3{}^2 + n_4{}^2) + g_{33}n_5{}^2 n_6 \\
+ g_{34}(n_1{}^3 - 3n_1 n_2{}^2 + n_3{}^3 - 3n_3 n_4{}^2) + g_{35}n_5(n_1{}^2 + n_2{}^2 - n_3{}^2 - n_4{}^2) \\
+ g_{41}n_6{}^4 + g_{42}n_6{}^2(n_1{}^2 + n_2{}^2 + n_3{}^2 + n_4{}^2) \\
+ g_{43}[(n_1{}^2 + n_2{}^2)^2 + (n_3{}^2 + n_4{}^2)^2 + 2n_1{}^2(n_3{}^2 + n_4{}^2) \\
+ 2n_2{}^2(n_3{}^2 + n_4{}^2)] + g_{44}n_5{}^2 n_6{}^2 + g_{45}n_5{}^2(n_1{}^2 + n_2{}^2 + n_3{}^2 + n_4{}^2) \\
+ g_{46}n_5{}^4 + g_{47}n_6[n_1(n_1{}^2 - 3n_2{}^2) + n_3(n_3{}^2 - 3n_4{}^2)] \\
+ g_{48}n_5 n_6(n_1{}^2 + n_2{}^2 - n_3{}^2 - n_4{}^2) + g_{49}[(n_1{}^2 + n_2{}^2)^2 + (n_3{}^2 + n_4{}^2)^2] \\
+ g_{40}n_5[n_1(n_1{}^2 - 3n_2{}^2) + n_3(n_3{}^2 - 3n_4{}^2)], \quad (Eq.1)\end{aligned}$$

There are six order parameters in this Landau polynomial, the first four of which correspond to the trimerization of Mo atoms (the $K_3$ irreducible representation). Note that the actual displacement of atoms represented by these four order parameters can be defined arbitrarily as long as a linear combination of two parameters within two sets (i.e. {$n_1$, $n_2$}, {$n_3$, $n_4$}) can each form three trimerized structures. Here, we define $n_1$ and $n_3$ as displacements correspond directly to the trimerization of Mo atoms while $n_2$ and $n_4$ to be atomic displacements perpendicular to $n_1$ and $n_3$. The $n_5$ and $n_6$ represent the displacement of Te layers against the Mo layer: the $n_5$ is the so-called polar mode (irreducible representation $\Gamma_2^-$), and $n_6$ is the "breathing" mode which changes the effective thickness of the structure (irreducible representation $\Gamma_2^+$). Corresponding atomic displacements of these order parameters are shown in Fig. 2(a).

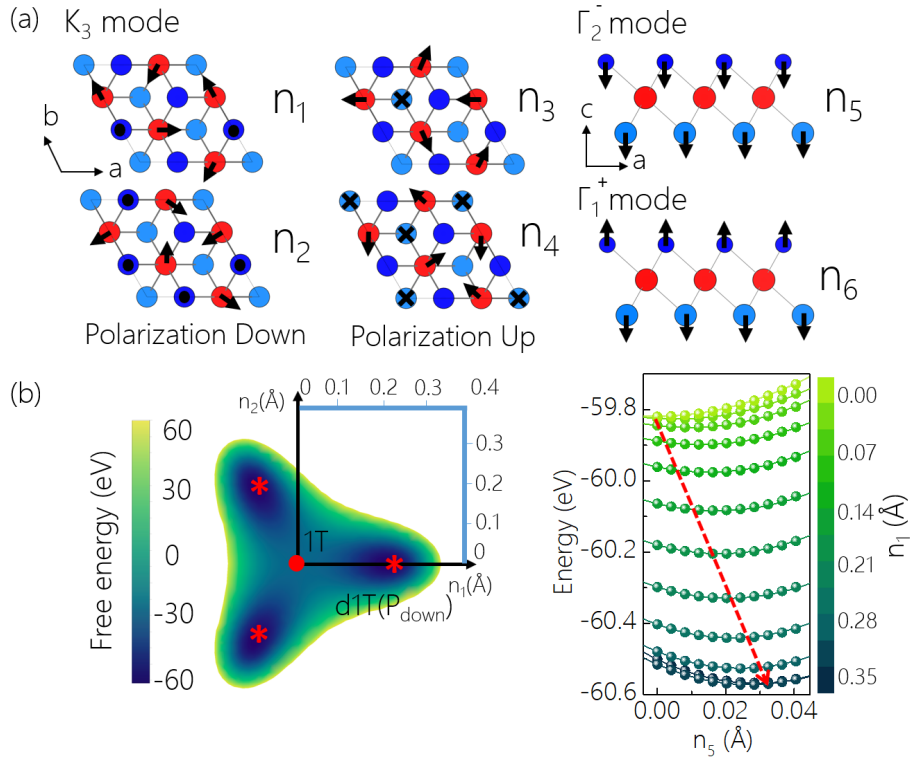

FIG 2 (a) Atomic displacement (black dot and cross stands for out and in-plane movement) of order parameters $n_1$, $n_2$, $n_3$, $n_4$, $n_5$ and $n_6$. $n_1$, $n_2$, $n_3$ and $n_4$ belongs to irreducible representation $K_3$ while $n_1$ and $n_2$ forms down polarized state and $n_3$ and $n_4$ forms up polarized state. (b) Left panel: down polarization plane formed by $n_1$ and $n_2$. Right panel: dependence of $n_5$ on $n_1$.

The most interesting thing about this free energy expression is that the first four order parameters $n_1$, $n_2$, $n_3$, $n_4$ are degenerate. Combining $n_1(n_3)$ and $n_2(n_4)$ linearly while suppressing $n_5$ and $n_6$, two polarization planes with up/down polarization can be formed [Fig. 2(b)]. The free energy expression of this polarization plane can be written as (taking $\{n_1, n_2\}$ as example):

$$\mathcal{F}(n_1, n_2) = g_{22}(n_1{}^2 + n_2{}^2) + g_{34}(n_1{}^3 - 3n_1 n_2{}^2) + (g_{43} + g_{49})[(n_1{}^2 + n_2{}^2)^2], \quad (Eq.\,2)$$

Each polarization plane has three symmetry related local minima corresponding to trimerized structures, bringing the total degeneracy of polarized state to 6 (3 polarization up structures and 3 polarization down structures). This degeneracy of the polarized structures can be easily understood: each Mo atom is surrounded by 6 Te atoms, trimerizing around three upper plane Te atoms will form three down polarized structures while trimerizing around 3 lower Te atoms will form 3 up polarized structures. Intuitively, one might assume the prototypical 1T phase is the only centro-symmetric structure linking the polarized d1T states. It is indeed the case when linking 6 degenerate d1T structures all together. However, there is other lower energy symmetric structures as the transition state between two of the adjacent ferroelectric states. By suppressing the $n_2$, $n_4$, $n_5$ and $n_6$, we reform the free energy expression within the subspace of $n_1$ and $n_3$:

$$\mathcal{F}(n_1, n_3) = g_{22}(n_1^2 + n_3^2) + g_{34}(n_1^3 + n_3^3) + g_{43}(n_1^2 + n_3^2)^2 + g_{49}[n_1^4 + n_3^4], \quad (Eq.3)$$

Analytically solving the equation $(\nabla \mathcal{F}(n_1, n_3)) = 0$ gives four energy extrema[Fig. S9(a)], two of which corresponds to two degenerate trimerized states ($\nabla \cdot (\nabla \mathcal{F}(n_1, n_3)) < 0$) the other one is the prototypical 1T structure where both $n_1$ and $n_3$ equals to zero ($\nabla \cdot (\nabla \mathcal{F}(n_1, n_3)) > 0$). The final one is a saddle point ($\nabla \cdot (\nabla \mathcal{F}(n_1, n_3)) = 0$) on the energy surface corresponding to a new distorted 1T structure with *C2/m* symmetry [insert in Fig. S3(b)]. This finding suggests the switching of ferroelectric state goes through this *C2/m* structure (t1T structure) rather than reversing back to the energy maxima 1T phase. Cycling through 2 of the total 6 degenerate d1T states, 6 degenerate t1T structures can be obtained. The total energy of this t1T structure is around 30meV/atoms lower than 1T structure, hence it should be more favorable when the system responds to the external electric field at finite temperature. To confirm this finding is indeed the global minimum, we also employed the transition state nudged elastic band (NEB) calculations, as shown in Fig. S3(b). Satisfyingly, the transition state calculation also suggests the t1T structure to be the best transition route of these ferroelectric switching.

To study the coupling relation between the polar mode $n_5$ and the trimerization mode $n_1$ ($n_2$, $n_3$, $n_4$), again, we rewrite the free energy expression with respect to $n_5$ and $n_1$:

$$\mathcal{F}(n_1, n_5) = g_{22}n_1^2 + g_{23}n_5^2 + g_{34}n_1^3 + g_{35}n_5n_1^2 + g_{45}n_5^2n_1^2 + g_{46}n_5^4 + (g_{49} + g_{43})n_1^4 + g_{40}n_5n_1^3, \quad (Eq.4)$$

Minimizing the free energy within this subspace while ignoring the high order terms (orders > 2), we get:

$$n_5 = -\frac{g_{35}}{2g_{23}} n_1^2, \quad (Eq.5)$$

As shown in the right panel of Fig. 2(b), the non-linear coupling between the primary order parameter $n_1$ and polar mode $n_5$ suggest the ferroelectricity rising from $n_1$ actually belongs to improper ferroelectric category. However, as will be presented below, comparing to the prototypical improper ferroelectrics like YMnO$_3$ where the electronic polarization rises form the polar mode[14], the ferroelectric dipole moment here originates solely from the major order parameters' ($n_1$, $n_2$, $n_3$, $n_4$) electronic contributions. To determine this electronic origin of ferroelectricity, we performed a detailed analysis from their electronic structures.

Typically, the finite spontaneous polarization for periodical system should be calculated using the modern theory of polarization. However, due to the vanishing charge density within the vacuum region, the charge center can be uniquely determined in the out-of-plane direction as a result of the broken periodical boundary conditions. A simple integration of the charge density can be used to determine the charge center in this system (at least in the non-periodical direction). The calculated polarization using this charge center method is 0.75×10$^{-12}$ C/m. To confirm this value, we also employed the conventional berry phase/Wannier charge center method. According to the modern theory of polarization, the calculation of ferroelectric polarization needs the dipole moment of two different structures: the centro-symmetric reference structure and the ferroelectric polar structure. However, since the non-periodicity in the out-of-plane direction, we can safely deem the centro-symmetric (1T or t1T phase) structure to have zero polarization value even if they are mantellic and beyond our reach with an ill-defined Berry phase. Our Berry phase/Wannier charge center method yields a final ferroelectric polarization as 0.88×10$^{-12}$ C/m, comparable to those in-plane polarized 2D ferroelectric materials like SnTe and SnSe[3, 4] while agree with the one obtained from charge center method. With previously defined order parameters, the polarization value is further decomposed into two contributions: the trimerization mode $n_1$ and the polar mode $n_5$. Results indicate the former one is much larger than the latter one ($1.61\times10^{-12}$ C/m vs $-0.85\times10^{-12}$ C/m), contradicting to the characters of the ordinary improper ferroelectricity[14]. This result confirms that the ferroelectricity in monolayer $AB_2$ structures is mainly from the electronic part of the polarization in the primary order parameter, while the second polar mode actually weakens the polarization, yielding a different behavior from the well-known improper ferroelectrics[14].

Furthermore, to examine the orbital contributions to the polarization value, a Wannier interpolation of Bloch bands was conducted. The projection functions were selected as a set of localized Hydrogen-like orbitals consisting of 1 $s$-orbital and 3 $p$-orbitals at each Te site and a $d_{xy}$ orbital at each Mo site. The Mo atoms' $d_{xy}$ orbitals are reoriented in order to achieve maximum localization and symmetry constraint (the orientation axes are shown in Table S3). From the decomposed orbital contribution in Table S9, the major contribution to the polarization does not come from the Mo atoms' orbitals. Instead, the polarization largely originates from the electronic orbitals of the Te atoms [Table S9]. Since in the defined order parameters $n_{1,2,3,4}$, the symmetry is dominated by the movement of Mo atoms, the origin of the polarization can only attribute to electronic part. To qualitatively confirm this, we calculated the polarization with and without the movement of Te atoms in $n_{1,2,3,4}$. The results show the polarization without Te's movement ($2.51\times10^{-12}$ C/m) is significantly larger than that with Te's movement ($1.61\times10^{-12}$ C/m), this negative contribution suggests the movement of Te atoms (~0.6Å) in $n_{1,2,3,4}$ also diminishes the total polarization and should originate from the depolarization field effect. As a result, this spontaneous polarization can only persist due to the large electronic contribution from Te atoms orbitals with the trimerization of the Mo atoms. This behavior should induce a drastically different electric field response than that of the traditional improper ferroelectric materials, which will be discussed later. On the other hand, as the polarization strongly depends on the non-vanishing first order parameter $n_1$, another question raises: *Why does AB$_2$ monolayers prefer d1T structure instead of reversing back to its centro-symmetric 1T phase?*

The electronic structure of ferroelectric materials has unique bonding character of the electrons. In displacive-type ferroelectrics, like PbTiO$_3$, the bonding contribution of Ti's *d*-orbitals to the neighboring Oxygen's *p*-orbitals are crucial toward lowering the total energy. To understand how the polar structure survives the depolarization field on a microscopic level, we closely examined the electronic structure against the structural transition with the help Wannier orbital overlap population method [1, 22]. Our results suggest a two-step process for the 1T phase transiting to the ferroelectric d1T state: the local Jahn-Teller type distortion followed by a hybridization between Mo atoms to form effective inter-site couplings.

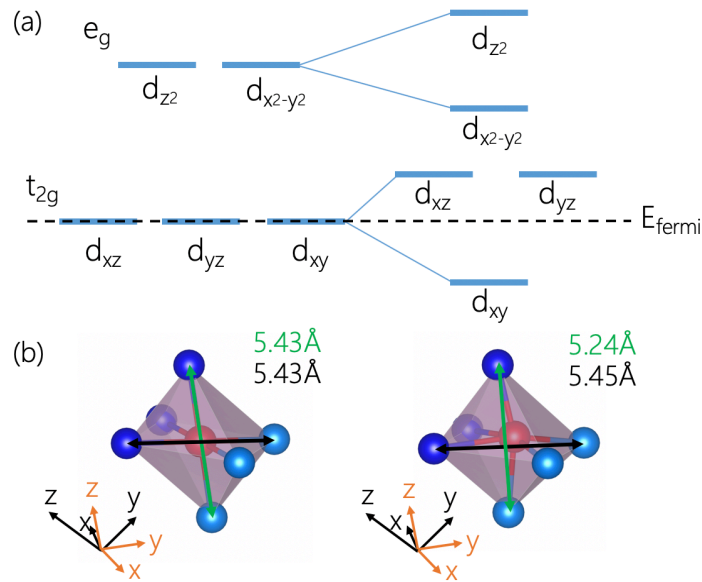

FIG 3 (a) Band diagram for Jahn-Teller like distortion within the Te octahedron. (b) Te octahedron of 1T (left) and d1T (right) showing the change of bond length and reorientation of axis, black arrow is the original axis.

Focusing on the Te octahedron in the centro-symmetric 1T phase as shown in Fig. 3, the Mo atoms are located at center of the crystal field, considering only nearest neighbor contribution to the symmetry, this structure yields two non-degenerate orbital sets: $e_g$ and $t_{2g}$, with two and three d-orbitals respectively(note that we reoriented the axis here). The lower energy $t_{2g}$ orbitals are three-fold degenerated and occupied by two electrons, thus the system is conductive. After the trimerization distortion, Mo atom moves along one of the three degenerate axes $Z_1$, $Z_2$ and $Z_3$ [Fig. 4(c)], and this movement breaks the symmetry of the crystal filed, further splits the energy levels into four subsets [right panel of Fig. 3(a)]. The $d_{xy}$ orbital becomes the lowest one and fully filled by two electrons, converting the system into insulating state. This Jahn-Teller like degeneracy breaking mechanism happens along with the change of bond length. As shown in Fig. 3(b), the length between upper and lower Te atoms decrease from 5.43 to 5.24 Å and the length between other two sets of Te atoms increase slightly from 5.43 to 5.45 Å.

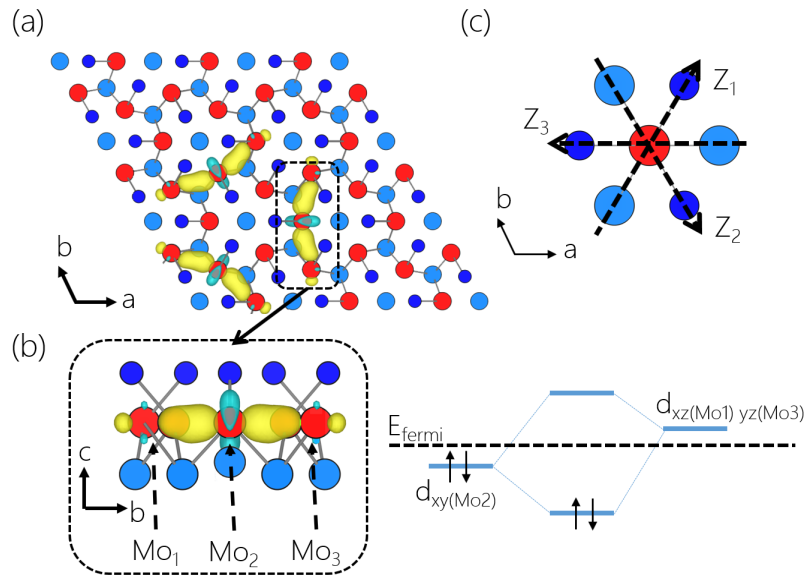

FIG 4 (a) Wannier function plot of three distinctively oriented valence Mo's $d_{xy}$ orbitals. (b) Left panel: Side view of one hybridized Mo's $d_{xy}$ orbital. Right panel: bonding diagram of the hybridization between Mo atoms' d-orbitals. (c) Three degenerate atomic displacement direction $Z_1$, $Z_2$ and $Z_3$ for the center Mo atoms.

Up until now, we have only focused on the distortions within a single Te octahedron, ignoring the ordering of Mo atoms' distortions (remember we have 3 degenerate distortion direction to choose from as shown in Fig. 4(c)). Without coupling between neighboring distortions, the system remains in a random configuration across the board. It turns out, the relation between neighboring octahedrons depends heavily on the hybridization of now occupied Mo's $d_{xy}$ orbitals. Indicated by the Wannier function plot [Fig. 4(a)], instead of localizing on the atomic site, the occupied Mo's $d_{xy}$ orbitals are spread out, showing traces of hybridization with neighboring Mo atoms' $d_{xy}$ and $d_{xz}$ orbitals. This three-site hybridization is crucial to the formation of d1T structure of AB$_2$ monolayers since it directly determines the coupling between neighboring cells. To quantify the covalency of this hybridization, the Wannier orbital overlap population (WOOP) analysis was adopted, decomposing the bonding orbitals with atomic-like localized orbital set (detailed methods can be found in the supporting materials[20]). To construct a complete set of orthonormal Wannier orbitals, projections were made on both valence and conduction bands to several localized orbitals. Using the "Valence+conduction" band generated Wannier orbitals as basis, the bonding orbitals (constructed from valence band only) are decomposed by projection onto the basis set, resulting in the overlap matrixes (projection coefficients). The decomposed charge number corresponding to the quasi-covalent $d_{xy}$ bond are shown in Table S2. Intuitively, the quasi-covalent bonding here has large contribution from the $d_{xz}$ and $d_{xy}$ orbitals of its neighboring Mo atoms. In order to quantitively measure the covalency, we define the covalency indicator as the charge percentage between different atomic wavefunctions:

$$CV = \frac{\sum Q_A}{\sum Q_B}, (Eq.6)$$

Where $CV$ stands for covalency. $Q_A$ and $Q_B$ represent the charge contribution from atom A and B, where atom B made the major contribution to the bond. For an ideal ionic bond, $CV$ should equals to 0, and, for an ideal covalent bond, equals to 1. For centro-symmetric PbTiO$_3$, WOOP analysis shows $CV$ equals to 0.17, suggesting the covalency between orbitals are sufficiently small and the bonding is largely ionic, this makes typical thin film of perovskite PbTiO$_3$ to be more vulnerable to the effect of depolarization field. However, in d1T MoTe$_2$ monolayers, $CV$ of the occupied $d_{xy}$ orbitals is 0.46, showing strong traces of covalency. Due to the directional properties of covalent bonding, the ordering of Mo's displacements is fixed by its bonding characters, giving rise to the anisotropic behavior of the imaginary phonon modes in Fig. 1. In other word, the hybridization of Mo's $d_{xy}$ orbitals further lowers the system's total energy, as indicated in the right panel of Fig. 4(b). These results indicate this in-plane quasi-covalent bonding is the origin of the trimerization of Mo atoms and it subsequently strengthened the ferroelectric phase against the depolarization field. It is also worth noting that for different A-site element (Cr, Mo, W), similar *CV* values [Table. S8] are obtained due to the changing of atomic radii. A smaller atomic radius results in a smaller lattice constant hence increases the covalency between A-site atoms. B-site elements (S, Se, Te) also affect the *CV* value: heavier B-site element increase the covalency while a lighter one decreases it.

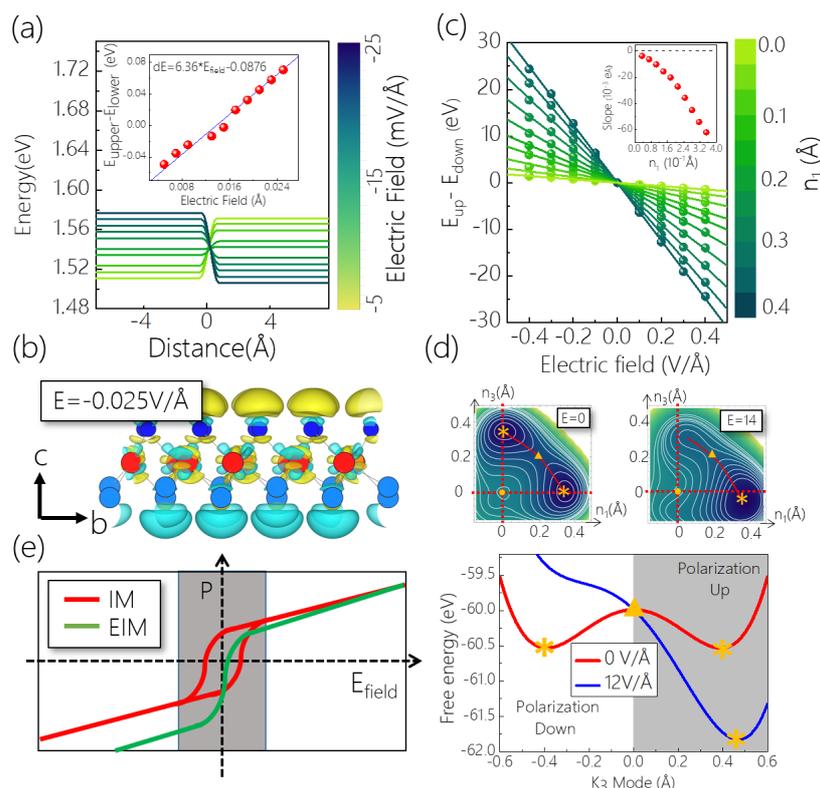

FIG 5 (a) Averaged local potential showing electronic response to the external electric field. Insert is the energy difference between left and right side of the local potential. The dipole correction layer is located at the periodic boundary(0Å) so a minus value corresponds to the upper vacuum space. (b) Differential charge density between d1T structure with and without electric field, yellow and blue indicate charge accumulation and depletion respectively. (c) Energy difference ($E_{up}$-$E_{down}$) vs external electric field strength. Insert shows the non-linear change of slop against the $n_1$ order parameter (d) Free energy contour with respect to different electric field strength, a star indicates a local minimum and a triangle represents a saddle point. (e) A sketch of dielectric response under low field for improper (IM) and this "electronic" improper (EIM) ferroelectricity in $AB_2$ monolayers. Note that there could still exist a very "slim" hysteresis in real life due to thermal fluctuation.

The most distinct difference of this materials from traditional improper ferroelectric materials is that the polarization largely arises from the electronic contribution of the primary order parameters instead of ionic part. This means the electronic polarization is more susceptible to the external electric field. On the other hand, decoupled polarization and structural order parameters makes the structure robust against it. To study the d1T phase's response to the electric field, an external electric field was applied along the vacuum direction of the d1T structures. At first, the dipole response to the external field is calculated [Fig. 5(c)]. During the process, the d1T structure doesn't change much if not at all while varying the electric field, we can see the dipole moment changes direction when electric filed strength reaches as low as -0.014 V/Å. This result shows the switching of the polarization's direction happens without the change of ionic positions at low field, indicating a weak coupling between geometry order parameters and the polarization. To study the geometry's response to the external electric field, a similar electric field with 0.4 V/Å (around 30 times higher than the one to revert the polarization direction without structural change) is applied to the asymmetrical slab. The geometry configuration was initially constrained to only allow $n_1$ trimerization order parameter to be active, eliminating the effect of polar mode $n_5$ and the breathing mode $n_6$ since they contribute negatively/significantly small towards the total polarization. By fixing the geometry while varying the electric field strength, the dependence of electric field against the polarization can be determined. As shown in Fig. S8(a), the optimum $n_1$ gradually deviates from the zero-field d1T structure when electric field is applied. Intuitively, with negative electric field, the upward polarization gradually moves towards down-polarized state while positive electric filed pushes the structure further towards upward polarized structures. The energy difference between two counterpart structures with opposite polarizations increases with applied electric field linearly [Fig. 5(b)]. However, the slop of this relation does not show linear dependency against the change of order parameter $n_1$. As shown in the insert of Fig.5(c), the slop shifted cubically with respect to the value of $n_1$ (As can be discerned from Fig. S8(b)). This suggests the coupling between different polarization structure to the electric field is non-linear: at low polarization state, the electric field has smaller impact towards the structure parameters while at high polarization its effect is more prominent. To include electric field effect in the Landau formula, an $E_{field}$ term is added to the total free energy expression:

$$\mathcal{F}(n_1, n_3, E_{field}) = \mathcal{F}(n_1, n_3) + g_{e0} \cdot E_{field} \cdot (g_{e1}(n_1 - n_3) + g_{e2}(n_1 - n_3)^3), \quad (Eq.6)$$

Note that due to the *unbound* effect of the electric field, system's total energy can always be lower by transferring valence electrons into conduction bands. To circumvent this conundrum, here the electric field term depends only on the energy difference between opposite polarized order parameters. By fitting this model to the DFT results, we estimated the electric field needed to transfer one state into another. As shown in Fig. 5(d), the energy shift with electric field and the non-favored structure vanishes at 12V/Å. This value is around 900 times higher than the one needed to reverse the polarization. To conclude above, the electric field needed to change the direction of the electrical polarization is significantly lower than the one to switch the geometrical structure. This means a shifted paraelectric like dielectric response (Fig. 5(e) green line) shall always exist in the low field region regardless of the frequency of the electric field. In ordinary displacive proper 2D out-of-plane ferroelectric materials like monolayer $In_2Se_3$, the polarization is strongly coupled to the order parameters, making the structure shift simultaneously with respect to the external electric field, yielding a hysteresis loop [Fig. 5(e) red line]. Our findings here can stimulate novel electronic devices based on this kind of material, such as both volatile random-access memories at low field and non-volatile random-access memories at high field. Combining the efficiency and non-volatile properties into one controllable device.

Another external field – the strain field also shows large manipulation against the energy barrier of transition t1T structure. As shown in Fig. S10, with biaxial strain applied (tensile or compressive), the ferroelectric d1T structure gains energy much more rapidly than the t1T structures. Applying either 14% tensile strain or 12% compressive strain, the system effectively loose it's ferroelectricity. Note that, since all calculations are done in 0K, the actual strain need to manipulating ferroelectricity could be drastically lowered with the help of thermal activation. Aside from the biaxial strain, the uniaxial strain will induce 1T' phase, breaking the ferroelectricity in the process. This degree of change against the ferroelectric structure gives more room for manipulation for monolayer MoTe$_2$ and could also be beneficial for industrial applications.

In summary, we report the origin of electronic-ionic decoupled improper ferroelectricity reside in distorted AB$_2$ monolayer structures. The Landau formula indicates the primary order parameters solely determines the symmetry of the ferroelectric d1T structures, different from ordinary improper ferroelectric materials in which secondary polar mode also help determines the final symmetry. A detailed bonding analysis shows the trimerizations between Mo atoms originates from the quasi-covalent bonding between neighboring Mo atoms and the polarization majorly originates from electronic part of this displacement. The electric field effect of the d1T structures confirms the difference between this ferroelectricity and ordinary improper ferroelectricity while the uniaxial strain gives sufficient degree of manipulation against the barrier between ferroelectric switching. With the development of electrical industry, these materials could be the key towards room-temperature or even high-temperature low-dimensional ferroelectric devices.

**Acknowledgements.**
X. L. thanks support from NSFC (No. 11804286) and the fundamental Research Funds for the central Universities. This work was supported by the National Natural Science Foundation of China (Grant No. 11974307 and 11774305), the National Key R&D Program of the MOST of China (Grant Nos. 2016YFA0300402), Zhejiang Provincial Natural Science Foundation (D19A040001), the Fundamental Research Funds for the Central Universities and the 2DMOST, Shenzhen Univ. (Grant No. 2018028). C. X would also like to thank Prof. Arash Mostofi and Dr. Nicholas Bristowe for their insightful suggestions.